\documentclass[12pt]{iopart}
\usepackage{graphicx}
\usepackage{upgreek}
\usepackage{color}

\begin{document}

\title{Ultraviolet laser spectroscopy of aluminum atoms in hollow-cathode lamp}

\author{Hongli Liu, Wenhao Yuan, Zetian Xu, Ke Deng and Zehuang Lu}

\address{MOE Key Laboratory of Fundamental Physical Quantities Measurement,

Hubei Key Laboratory of Gravitation and Quantum Physics, School of Physics,

Huazhong University of Science and Technology, Wuhan 430074, China
}
\ead{zehuanglu@hust.edu.cn}
\vspace{10pt}
\begin{indented}
\item[] February 2018
\end{indented}

\begin{abstract}
We report precision measurement of aluminum atoms ${^{2}P_{1/2}}-{^{2}S_{1/2}}$ transition at 394 nm and ${^{2}P_{3/2}}-{^{2}S_{1/2}}$ transition at 396 nm in a hollow-cathode lamp (HCL). Both absorption spectroscopy and saturated absorption spectroscopy (SAS) are performed. From the absorption spectroscopy the Doppler linewidth is estimated to be 2.6 GHz. The SAS spectroscopy is analyzed based on the velocity-changing-effect model. With a frequency comb calibrated wavemeter, the frequencies of  ${^{2}P_{1/2}},{F=3}-{^{2}S_{1/2}},{F=2}$ transition and  ${^{2}P_{3/2}},{F=4}-{^{2}S_{1/2}},{F=3}$ transition are measured to be 759.905401(10) THz and 756.547403(10) THz, respectively. The hyperfine structure constants of aluminum atoms are determined and compared with previously reported measurement results and theoretical calculation. Reasonable agreement is found for the magnetic dipole constant (A constant), while the electric quadrupole constant (B constant) has a large deviation.
\end{abstract}

%
\vspace{2pc}

%
%
%

\section{Introduction}
Aluminum atoms transitions from the $3s^23p$ ${^{2}P_{1/2} }$ and $ {^{2}P_{3/2}} $ levels to the $3s^24s$ ${^{2}S_{1/2}} $ level at 394 nm and 396 nm have been identified in stellar atmospheres \cite{01}. Precision spectroscopy of aluminum atoms has many applications in astrophysics. For example, non-local thermodynamic equilibrium (NLTE) modeling based on the hyperfine structure (HFS) plays a key role in determining the chemical composition in the Sun \cite{02}, in high resolution abundance analysis of subgiant \cite{04}, red giants \cite{05}, and late-type stars \cite{07}, and in studies of the chemical enrichment of the Galaxy and of globular clusters \cite{06}. In addition, precision spectroscopy of aluminum atoms also plays an important role in the studies of the variation of fundamental constants with time and space in an expanding Universe \cite{08,09,10,11}.

The UV laser spectroscopy of aluminum atoms is also important for resonance-enhanced photoionization of aluminum atoms, which will serve as a high efficiency, low contamination way to load aluminum ions for Al$^+$ ion optical clocks \cite{12}. Al$^+$ ion optical clocks have low sensitivity to electromagnetic perturbations, narrow natural linewidth \cite{13} and the smallest sensitivity to blackbody radiation \cite{14}. It shall be noted that resonance-enhanced photoionization is not jut limited to aluminum atoms, but has been demonstrated for many other elements as efficient ways to load singly charged ions \cite{15,16,17,18}.

\begin{figure}
  \centering
  \includegraphics[width=12 cm]{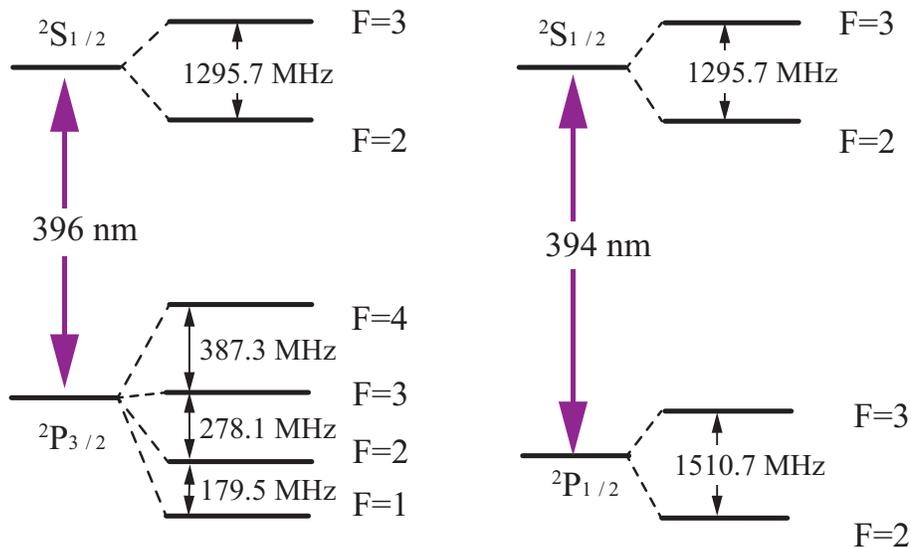}\\
  \caption{The hyperfine energy levels of Al}\label{fig:1}
\end{figure}

The first excited state of aluminum atoms is ${^{2}S_{1/2}}$  and the ground states of aluminum atoms have two fine-splitting energy levels  ${^{2}P_{1/2}}$ and $ {^{2}P_{3/2}}$. The aluminum atoms have almost equal population distributions of the two ground states after emitting from hot ovens \cite{19}. There are four hyperfine transitions for the ${^{2}P_{1/2}}- {^{2}S_{1/2}}$ transition at 394 nm and six hyperfine transitions for ${^{2}P_{3/2}}- {^{2}S_{1/2}}$ transition at 396 nm, as shown in \Fref{fig:1}. We will use short-hand notation for these transitions throughout the paper. For example, for the 394 nm transition, the ${^{2}P_{1/2}},{F=3}- {^{2}S_{1/2}},{F=2}$ transition is short for 3-2 transition and cross-over transition  ${^{2}P_{1/2}},{F=3} - {^{2}S_{1/2}},{F=2,3}$ is short for {c.o.3-2,3} transition
 etc., and for the 396 nm transition, the ${^{2}P_{3/2}},{F=4}- {^{2}S_{1/2}},{F=3}$ transition is short for 4-3 transition, etc.. The transition rate for the 394 nm transition is
$4.99\times10^7$ s$^{-1}$ and the saturation intensity is $I_{sat}=16.9 $  mW/cm$^2$. The transition rate and the saturation intensity for the 396 nm transition is $9.85\times10^7$ s$^{-1}$ and 32.9 mW/cm$^2$.

The aluminum atom ground-state hyperfine constants have been measured with atomic beam technology before \cite{20}, but the B hyperfine constant of
$^{3}P_{3/2}$ level is different with the result obtained with far-infrared laser magnetic resonance method \cite{21}. In addition, there is a large difference between experimental data and theoretical calculations \cite{22}. Consequently, it is worthwhile to perform high precision measurement of aluminum hyperfine constants.

In most atom optics experiments, the atomic transition is observed in a vapor cell or in an atomic beam. In a regular vapor cell, producing sufficient atomic gases of aluminum requires a temperature in the order of 1500 K, which is very difficult to realize, and atomic beam technology is much more complicated and expensive. As an alternative, we used a hollow cathode lamp (HCL), in which a high vapor pressure is realized by sputtering of the noble gas. They are usually used to perform high-resolution spectroscopy of metal vapors of which the vapor pressure is very low at room temperature \cite{23}, and have been used to observe the transitions for Cd \cite{24,25}, Sr \cite{26}, Ca \cite{27}, Yb \cite{23}, Yb$^+ $ \cite{28}, Fe \cite{29} and Tl \cite{30}.

Here we investigate the use of HCL as an efficient way to produce a high-density vapor of aluminum atoms for saturated absorption spectroscopy (SAS). Because the UV laser at 394 nm and 396 nm cannot be measured directly by the optical frequency comb in our lab, a high resolution wavemeter (HighFinesse WS-U1021) is used to measure the frequency of the UV lasers, and the wavemeter is calibrated by the second harmonic of an ultra-stable laser whose frequency is measured by the frequency comb. The absorption spectra are measured to analyze the optical density (OD) and Doppler linewidth in the HCL. The saturated spectroscopy with the velocity changing collision (VCC) was considered for the calculation of the hyperfine transitions of aluminum atoms.

\section{Experimental Setup}

\begin{figure}
  \centering
  \includegraphics[width=12 cm]{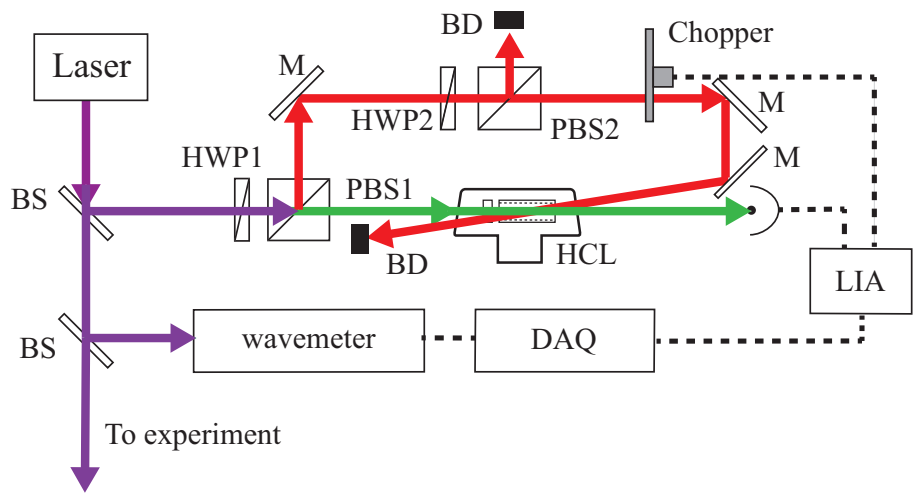}\\
  \caption{Schematic of experimental setup. BS: beam splitter; HCL, hollow cathode lamp; PBS, polarization beam splitter; HWP, half-waveplate; QWP, quarter-waveplate; M, mirror; PD, photodiode; BD, beam dump, LIA: lock-in amplifier; DAQ: data acquisition system. The red line shows the pump laser path and the green line represents the probe laser path.}\label{fig:2}
\end{figure}

The schematic of the experimental setup is shown in \Fref{fig:2}. Two ultraviolet external-cavity-diode-lasers (Toptica DL pro) at 394 nm and 396 nm are used as the laser sources. Coarse adjustment of the wavelengths of the lasers can be achieved by manually adjusting the grating positions of the lasers, and fine adjustment of the wavelengths is realized by tuning piezoelectric transducers that are glued to the gratings. The laser output is split into three parts: one is for laser spectroscopy, one is sent to the wavemeter, and the last part is sent to ion trap experiment.

The uncertainty of the wavemeter is better than 0.3 MHz at measurement cycle of dozens seconds, which is mainly limited by temperature fluctuations in the lab \cite{31}. The wavemeter is calibrated by a 534 nm laser which is generated by frequency doubling of an ultra-stable laser at 1068 nm. The ultra-stable laser has a frequency instability of $5.0 \times 10^{-16} $ and a frequency drift rate of 20 mHz/s \cite{32}, and the frequency of the ultra-stable laser is measured by a frequency comb that is referenced to a Cs clock (Symmetricom 5071A). The calibration wavelength is chosen at 534 nm instead of 1068 nm so that the calibration wavelength is closer to the measurement wavelengths. It is estimated that the inaccuracy introduced by this calibration process is less than 10 MHz.

The powers of the pump laser and the probe laser are adjusted by two half waveplates and PBS. For the 394 nm laser, at the center of the HCL the probe beam has an intensity 1/e$^2$ radius of 0.4 mm horizontally and 0.7 mm vertically, and the pump beam has intensity 1/e$^2$ radius of 0.9 mm horizontally and 0.8 mm vertically. As for 396 nm, the diameter of the probe beam and pump beam changes to 0.4 (0.4) mm and 1.1 (0.9) mm horizontally (vertically), respectively. A mechanical chopper in the pump beam path is introduced to improve the signal to noise ratio through lock-in detection. The lock-in amplifier that we used is SR830 from Stanford Research Systems. The frequency of the laser and the detected signal of the HCL is acquired through the data acquisition system.

The HCL (Hamamatsu L2783-70NE-Al) is constructed with a see-through cathode, a ring-shaped anode mounted inside a T-shaped Brewster glass bulb, and is filled with neon buffer gas. The spatial distribution and the collision effect of the target atoms is inhomogeneous in the see-through cathode region \cite{29}.To reduce the collision effect, the laser is made to pass through the lamp at the center of the cathode.

\section{Results and Analysis}

\subsection{The absorption spectroscopy}

\begin{figure}
  \centering
  \includegraphics[width=12 cm]{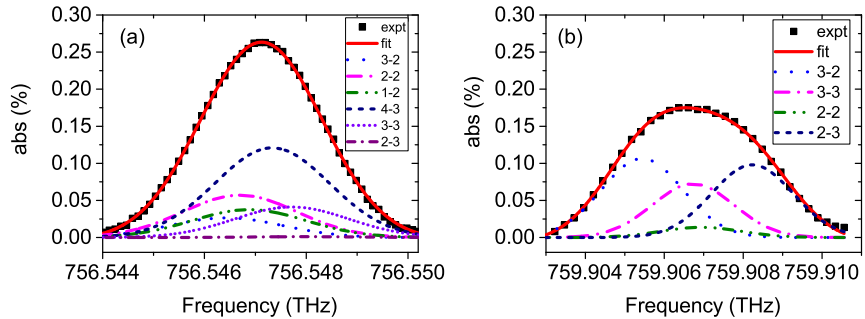}\\
  \caption{ The absorption spectra of Al in a HCL. (a) the absorption spectra for the 396 nm transition. The contribution of the six individual hyperfine transitions are also shown. (b) the absorption spectra for the 394 nm transition. The contribution of the four individual hyperfine transitions are also shown.}\label{fig:3}
\end{figure}

\Fref{fig:3} shows the absorption spectra of the Al. The vertical axis in the figure is the relative absorption of the HCL. The absorption spectra are obtained at a probe beam power of 3  $\upmu$W and the HCL current is set at the maximum current of 20 mA. From the spectra, we can see that the maximum absorption is at 756.547 THz for the 396 nm transition and the full width at half maximum (FWHM) linewidth is 2.8 GHz. As for the 394 nm transition, the maximum absorption is at 759.907 THz and the FWHM linewidth is 4.4 GHz.

\begin{figure}
  \centering
  \includegraphics[width=12 cm]{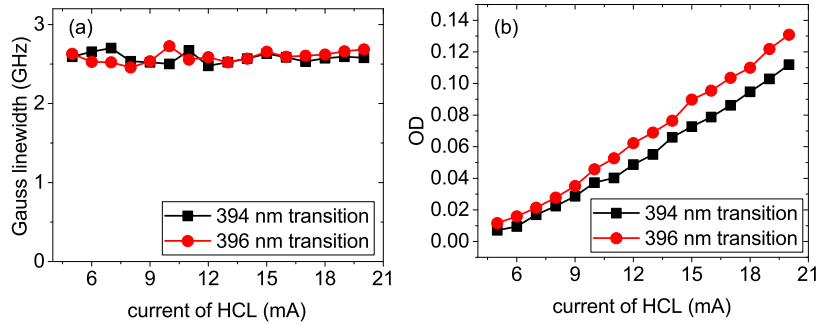}\\
  \caption{Dependence of the absorption spectra on the HCL discharge current. (a) Dependence of Doppler linewidth on the HCL discharge current. (b) Dependence of OD on the HCL discharge current.}\label{fig:4}
\end{figure}

Because of line overlapping, the real linewidth caused by Doppler broadening and absorption for the hyperfine transitions is determined by multiline curve fitting. From the fitting, the Doppler linewidth was estimated to be 2.6 GHz for both 396 nm and 394 nm transitions, which corresponds to effective temperature of 620 K. The maximum relative absorption of 394 nm hyperfine transition is 0.11, corresponding to an optical density (OD) of 0.11, where the OD is defined as $\textrm{OD}=-\textrm{ln}(1-\textrm{abs}) $. The maximum OD for the 396 nm transition is estimated to be 0.13.

The relationship of the measured OD and Doppler linewidth with the HCL discharge currents are shown in \Fref{fig:4}. We can find that there is no obvious dependence of the Doppler linewidth with the discharge current, which indicate that the effective temperature of the aluminum atoms do not change with the discharge current. The OD increases with the discharge current without saturation effect, and the maximum absorption is obtained at the maximum current of 20 mA.

\subsection{The saturated absorption spectroscopy}

\begin{figure}
  \centering
  \includegraphics[width=12 cm]{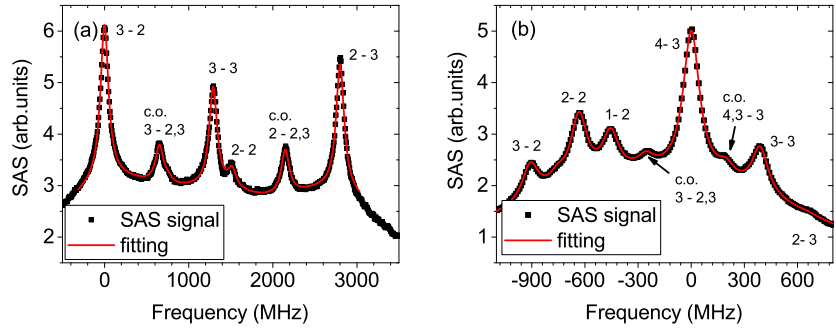}\\
  \caption{The SAS of 27Al at center of the HCL. (a) The SAS for 394 nm transitions, (b) The SAS for 396 nm transitions.}\label{fig:5}
\end{figure}

The SAS is observed with a probe power of 3 $\upmu$W and a pump power of 2 mW (1.6 mW) for the transition at 394 nm (396 nm). The observed spectra are shown in \Fref{fig:5}. For the 394 nm spectrum, the strongest hyperfine transition is 3-2 transition. All four hyperfine transitions can be observed. In addition, two cross-over signals which correspond to the {c.o.2-2,3}   and {c.o.3-2,3} are observed. As for the 396 nm spectrum, the strongest hyperfine transition is 4-3 transition. A little peak at the edge of the high frequency end of the spectrum corresponds to the weak 3-2 transition. Two cross-over transitions,  {c.o.3-2,3} and {c.o.4,3-3},  are also observed.

There are Doppler pedestals in the SAS spectra, which are due to the velocity changing collisions. For strong collision, the SAS profile can be expressed as \cite{33}

\begin{eqnarray}\label{eq1}
   S(\nu)=k\{L(\nu,\Delta\nu_H)+C\exp[-(\nu/\Delta\nu_{D})^2]\}\times\frac{\exp[-(\nu/\Delta\nu_{D})^2]}{1+C\exp[-(\nu/\Delta\nu_{D})^2]},
\end{eqnarray}
where $L(\nu,\Delta\nu_H) $ is the Lorentz function with the homogeneous broadening linewidth of $\Delta\nu_H $, $\Delta\nu_{D}$ is the Doppler broadening linewidth. $k$ is an intensity scale factor, which indicates the transition strength, and $C$ gives the weight of the unwanted VCC effect \cite{34}.

The parameters of the transition strength $k$, Lorentz linewidth $\Delta\nu_H $ and VCC parameter $C$ can be determined by least-squares-fitting of the measured spectra. Since the hyperfine transition frequencies of aluminum atoms are very close, with the closest transition having a frequency separation of only 179.8 MHz, all the transition lines will partially overlap. This will shift the peak of the SAS signals. Taking into account the overlap of the transition lines, the measured SAS signal can be expressed as
\begin{equation}\label{eq2}
  y(\nu)=y_0+\sum a_iS(\nu-\nu_i),
\end{equation}
where $y_0$ is the offset of the signal and $a_i$ is the relative transition intensity, $ \nu_i$ is the center of the corresponding hyperfine transition.

The experimental results and curve fitting results are shown in \Fref{fig:5}. For the curve fitting process, the Lorentz linewidth of the observed hyperfine transitions are assumed to be the same.

\begin{table}
\caption{\label{tab2}The experiment(expt) and calculated(cal) relative line intensity(Rel. int) and the line center offset of 394 nm transitions.}

\begin{indented}
\lineup
\item[]\begin{tabular}{@{}*{4}{l}}
\br
Transition&Rel. int&Rel. int&line center offset\cr

F-F'&expt&cal&(MHz)\cr
\mr
3-2&$1$&1&0\cr
3-3&$0.64$&0.8&1292.9(0.4)$^{\rm a}$\cr
2-2&$0.12$&0.2&1519.6(2.6)\cr
2-3&$0.81$&1&2802.7(0.4)\cr

\br
\end{tabular}
\item[] $^{\rm a}$ the value in the brackets are the standard error  of the measurement in 5 measurements.
\end{indented}
\end{table}

\begin{table}
\caption{\label{tab3}The experiment(expt) and calculated(cal) relative line intensity(Rel. int) and the line center offset of 396 nm transitions.}

\begin{indented}
\lineup
\item[]\begin{tabular}{@{}*{4}{l}}
\br
Transition&Rel. int&Rel. int&line center offset\cr

F-F'&expt&cal&(MHz)\cr
\mr
3-2&$0.23$&0.4&-909.3(0.2)$^{\rm a}$\cr
2-2&$0.48$&0.3&-633.4(0.1)\cr
1-2&$0.32$&0.4&-452.0(0.2)\cr
4-3&$1$&1&0\cr
3-3&$0.34$&0.7&388.6(0.1)\cr
2-3&$0.01$&0.1&669.4(9.0)\cr
\br
\end{tabular}
\item[] $^{\rm a}$ the value in the brackets are the standard error  of the measurement in 16 measurement.
\end{indented}
\end{table}

\begin{table*}
\caption{\label{tab1}The hyperfine constant A and B of aluminum atoms.}

\begin{indented}
\lineup
\item[]\begin{tabular}{@{}*{5}{l}}
\br
Parameter&Present work&Previous values$^{\rm a}$&Previous values$^{\rm b}$&Theoretical values$^{\rm c}$\cr
\mr
A($^2S_{1/2}$)(MHz)&\m $431.89(11)$&\m 431.84(91)&\m ---&\m 407.18\cr
A($^2P_{1/2}$)(MHz)&\m $503.58(27)$&\m 502.04(97)&\m 502.0346(5)&\m 498.33\cr
A($^2P_{3/2}$)(MHz)&\m $94.47(8)$&\m 93.76(71)&\m 94.27723(10)&\m 100.98\cr
B($^2P_{3/2}$)(MHz)&\m $11.75(37)$&\m 19.12(86)&\m 9.45949(35)&\m 19.59\cr

\br
\end{tabular}
\item[] $^{\rm a}$ Values measured with atomic beam technology \cite{20}.
\item[] $^{\rm b}$ Values measured with far-infrared laser magnetic resonance method \cite{21}.
\item[] $^{\rm c}$ Theoretical values calculated with unitary coupled cluster singles and double (UCCSD) mehods \cite{22}.
\end{indented}
\end{table*}

From the curve fitting, the homogeneous broadening linewidths $\Delta\nu_H $ are determined to be 116 MHz and 111 MHz for the 394 nm transition and 396 nm transition, respectively. The homogeneous broadening has contributions from the lifetime broadening, the pressure broadening, power broadening and  transit-time broadening. The transit-time broadening is estimated to be 19 kHz, which has negligible contribution to the homogenous broadening. The homogenous broadening linewidth of the SAS signal with combination of lifetime, pressure, and power broadening can be described by the following expression \cite{35}

\begin{equation}\label{eq3}
  \Delta\nu_H=\Delta\nu_n\sqrt{\frac{I}{I_{sat}}+(1+\frac{\Delta\nu_p}{\Delta\nu_n})^2},
\end{equation}
where $ \Delta\nu_p$ is the pressure broadening linewidth and $\Delta\nu_n$ is the lifetime broadening linewidth. The pressure broadening linewidth is estimated to be 107 MHz for 394 nm and 93 MHz for 396 nm transition.

The fitted relative line intensity is listed in \Tref{tab2} and \Tref{tab3}. The fitted line intensity is used in the absorption fitting in Section 3.1. We find that the fitted relative line intensity is not consistent with the theoretical value calculated with Clebsch-Gordan coefficients. This is due to the fact that the energy level distribution was changed by the complicated collision processes in the discharge of the HCL.

The line center offsets of the hyperfine transitions relative to the corresponding baseline are determined from the fitting, and the standard error of the measurement is listed in \Tref{tab2} and \Tref{tab3}. The main error of the line center offset comes from the uncertainty of the wavemeter and is estimated to be 0.3 MHz.
The frequency of the two baseline transitions of ${^{2}P_{1/2}},{F=3}- {^{2}S_{1/2}},{F=2}$ and ${^{2}P_{3/2}},{F=4}- {^{2}S_{1/2}},{F=3}$ are measured to be 759.905401(10) THz and 756.547403(10) THz, respectively. The uncertainty of the absolute frequency is mainly limited by the accuracy of the wavemeter. The pressure shift is not considered here and will be studied in the future by changing the pressure of the HCL.

The hyperfine constants A and B are estimated with the fitted SAS spectra at 394 nm and 396 nm. In the calculation of the hyperfine constants, all the observed hyperfine transitions are used with weighted averaging.
The results are listed in the \Tref{tab1}.
The uncertainty of the experimental data depends on the accuracy of the relative frequency difference between hyperfine transitions, and is mainly due to three sources:
1) the uncertainty of the wavemeter of 0.3 MHz.
2) the statistical error of the experiments. For the ${^{2}P_{1/2}}$ to the $^{2}S_{1/2}$, the standard error of the measured resonant transition is 0.4 MHz for the 3-3 and 2-3 transitions and 2.6 MHz for the 2-2 hyperfine transition. As for the $^{2}P_{3/2}$ to the ${^{2}S_{1/2}}$ transition, less than 0.3 MHz standard error is used for most of the hyperfine transitions, except the 2-3 transition which has a standard error of 9 MHz because of weak transition intensity.
3) the pressure shift induced by the HCL. Although the absolute frequency of the transition will mainly be influenced by the pressure shift, but the relative frequency difference between hyperfine transitions will not be adversely affected if
we assume all the hyperfine energy levels have similar pressure shift coefficients.

The uncertainty of the magnetic
dipole constant A is 0.11 MHz for the state  of   ${^{2}S_{1/2}}$, and 0.27 MHz for the  $^{2}P_{3/2}$ . As for the  ${^{2}P_{3/2}}$  energy level, the uncertainty of the magnetic dipole constant A is estimated to be 0.08 MHz, and the uncertainty of the electric quadrupole constant B is estimated to be 0.37 MHz. From \Tref{tab1} we can see that A constants agree with the previous measurements very well. However, the B constant of the ${^{2}P_{3/2}}$ state has a large difference among different experiments. The exact reason is worthy of further investigation.

\section{Conclusion}

In conclusion, we have performed precision spectroscopy of aluminum atoms at 394 nm and 396 nm in a HCL. The frequencies of the transitions are measured with a wavemeter which is calibrated by the second harmonic of an ultra-stable laser, whose frequency is measured by an optical frequency comb.

The Doppler linewidth was estimated to be 2.6 GHz and the maximum OD of the 394 nm transition and 396 nm transition are estimated to be 0.11 and 0.13, respectively. The SAS of the aluminum atoms in HCL is recorded. Four hyperfine transitions and two cross-over signals are observed in the SAS of 394 nm, and six hyperfine transitions and two cross-over signals are observed for the 396 nm spectroscopy.

The VCC effect is considered to analyze the SAS result. The hyperfine transition frequencies are estimated by curve fitting, and the frequencies of ${^2P_{1/2}},{F=3}-{^2S_{1/2}},{F=2}$ and ${^2P_{3/2}},{F=4}-{^2S_{1/2}},{F=3}$ are found to be 759.905401(10) THz and 756.547403(10) THz, respectively. The A constant of the ${^2S_{1/2}}$ and ${^2P_{1/2}}$ are calculated to be 431.89 and 503.58 MHz, respectively.  A constant and B constant  for the ${^2P_{3/2}}$ state are calculated to be 94.47 and 11.75 MHz, respectively. The uncertainty of the hyperfine constant A are below 0.3 MHz.

Since the pressure of the HCL is unknown, in the future we will measure the pressure broadening and pressure shift by changing the pressure in the HCL. In addition, frequency locking with the polarization spectroscopy technique will be investigated too.

\section*{\ack}
The project is partially supported by the National Key R\&D Program of China (Grant {No.2017YFA0304401}) and
the National Natural Science Foundation of China (Grants {No.11304109}, {No.11774108}, and {No.91336213}).

\end{document}